# Distributed Web browsing: supporting frequent uses and opportunistic requirements

Sergio Firmenich[1,2]  · Gabriela Bosetti[1] · Gustavo Rossi[1,2] · Marco Winckler[3] · José María Corletto[1]

**Abstract**
Nowadays, the development of Web applications supporting distributed user interfaces (DUI) is straightforward. However, it is still hard to find Web sites supporting this kind of user interaction. Although studies on this field have demonstrated that DUI would improve the user experience, users are not massively empowered to manage these kinds of interactions. In this setting, we propose to move the responsibility of distributing both the UI and user interaction, from the application (a Web application) to the client (the Web browser), giving also rise to inter-application interaction distribution. This paper presents a platform for client-side DUI, built on the foundations of Web augmentation and End User Development. The idea is to empower end users to apply an augmentation layer over existing Web applications, considering both frequent use and opportunistic DUI requirements. In this work, we present the architecture and a prototype tool supporting this approach and illustrate the incorporation of some DUI features through case studies.

## 1 Introduction

The distribution of user interfaces (DUI) was a growing trend during the last 10 years. Even earlier, by the mid-nineties, the first concepts of DUI appeared [6] with the advent of migratory applications [2]. DUI research has gained relevance and it advanced to really complex systems including several dimensions: how input and output are managed, the device in which the application runs, the context where the application is used, when it is used, etc. [6]. Other works also addressed collaboration as another dimension of DUI systems [16].

* Sergio Firmenich
  sergio.firmenich@lifia.info.unlp.edu.ar

  Gabriela Bosetti
  gabriela.bosetti@lifia.info.unlp.edu.ar

  Gustavo Rossi
  gustavo@lifia.info.unlp.edu.ar

  Marco Winckler
  winckler@irit.fr

1 LIFIA, Facultad de Informátca, UNLP, La Plata, Argentina
2 CONICET, La Plata, Argentina
3 ICS-IRIT, University of Toulouse 3, Toulouse, France

According to Vanderdonckt [23], part of the need of DUI comes from the fact that "*a user is rarely using one single platform at a time, but several different platforms at a time or one after another, and a user is no longer staying in the same environment since she is moving from one environment to another or across environments*". Although this definition was stated in 2011, so far, this is not only still valid but also more relevant given that the same aspects are nowadays magnified due to the continuous emergence of new types of Web applications and mobile devices.

In the context of Web applications, there are approaches such as Panelrama [26] and Connichiwa [22] that provide a clear support for developing Web applications supporting DUI from scratch. Recently, some work [3] focused on the *opportunistic* distribution based on gestures, by providing another framework for developing cross-device Web application interactions. Among others with similar goals (as we comment in the Related Works section), these are important contributions to the field. However, most Web applications do not offer this kind of interaction, even in spite of knowing how DUI behaviour could improve the user experience.

In this paper, we extend our previous research [7] on empowering end users with the possibility to apply DUI features on existing Web applications, and not necessarily the ones foreseen by a developer. When Web applications do not offer those features that users may need, experience has shown that the crowd of users react trying to satisfy these needs by themselves. This is a very common practice in Web Browsing Augmentation, i.e. using tools (usually deployed as Web Browser Extensions) to augment Web application capabilities. To cite one example, a simple solution for cross-device interaction called "Slides—Presentation Remote"[1] has more than sixty thousand users, offering a remote control for presentations in some well-known Web applications (Google Drive, SlideShare, Prezi, etc.). Examples like this one—and thousand others—clearly show that while ad-hoc developers may create applications providing this kind of web augmentation layer, there are users expecting them, and we want to empower them to be the ones that produce their solutions, no matter their technical knowledge.

Web Browser Augmentation, as a technique, is a perfect target for end-user-driven DUI applications, such as supporting opportunistic or frequent layout or interaction distribution and UI migration. For instance, consider a scenario where an initial student is starting to learn about Web development in jsfiddle.net. He has a netbook with a small screen, so it is not pleasant to use such interface in such device. The main area in the UI of such application is originally split in four pieces: three input boxes (JS, HTML, CSS) and a preview one. He may want to open a second tab in his browser and split such User Interface (UI) by distributing the CSS and preview components to the second tab, so he can write the HTML and JS code in a comfortable way, without visualizing it with line breaks. The desired DUI feature is not originally supported by jsfiddle.net, but a third-party tool can add it through Web Augmentation.

The aforementioned situation may represent an opportunistic scenario for a user who hardly uses a device of small dimensions for programming, but also a frequent use scenario for those who only have access to a notebook, or the ones that are used to a second screen for their daily life activities. Both modalities—opportunistic and frequent—are addressed in our approach in two different ways. The first one is about allowing the end user to distribute an application on demand; he points out which DOM element he wants to distribute, so it can be collected by the tool and distributed when he explicitly asks to move it to another context (e.g. another device, tab, window, etc.). A second modality is allowing the end user to define a context rule conditioning the distribution, and performing the distribution automatically, when the rule is successfully evaluated.

The combination of DUI and Web augmentation also contributes to the discussion about cross-application DUI, i.e. how user interfaces from two Web Sites could react when the user is interacting with each other from distinct devices/displays. Consider a workstation with two displays. While in one of them the user may be reading a Wikipedia article, the second one could be displaying a carousel with related images from Google Images. This interaction could make us conceive it as a mashup application; however, it is not, since it does not integrate existing application's content into a new one, but coordinating the UI of two different applications under the user's requirements when he is using more than one display (a tab, a single window) in the browser. This would be possible because Web augmentation artefacts usually run as part of the Web browser, so they can be aware of every Web site loaded. In this way it is possible to remove the boundaries of a single application. Using some Web augmentation techniques, cross-applications and cross-devices DUI interactions could be easily applied by end users with the correct set of abstraction tools.

We strongly consider that improving how users access Web content and Web sites information is a crucial aspect, and nowadays, it is very important to support this content access from a multi devices point of view. In this work, we propose an end-user-driven architecture, where users are the ones who apply DUI features and behaviours. In addition, we provide a framework to allow advanced users or programmers to define new DUI-oriented behaviour, following the philosophy behind the Web-customization communities, where end users and developers coexist and collaborate altogether.

This paper presents the approach, the overall architecture and the supporting tools through some case studies. It is structured as follows. First, in Sect. 2, we provide the underlying background. Section 3 introduces our approach and the main components of the underlying architecture. The supporting tools are illustrated via case studies in Sect. 4. Related works are presented and compared in Sect. 5. In Sect. 6, we analyse other DUI dimensions that could be tackled with Web augmentation. Finally, we discuss our contribution in Sect. 7, in conjunction with our future work.

## 2 Background

The Web has become a widely customizable space, making it more specific to the needs of each individual user. Personal Information Management (PIM) systems allow end users to collect and reorganize information according to their interests or activities, while Web Augmentation (WA), in conjunction with End-User Programming [18, 20] techniques, allow them to enhance the Web by manipulating its content according to their requirements, for building a solution on the basis of what already exists. Below, we introduce the concepts we use as a basis for building DUI in Web applications.

---
[1] https://chrome.google.com/webstore/detail/slides-presentation-remot/mhfdnafbhfglkcjgkgoopjoadaopcomi.

## 2.1 Personal Information Management systems and the Web

People store information with a purpose; they pick it from different sources, put it together, choose an order and a place, with the purpose of having it available for further making inquiries. The same process happens in the real and digital world, but through different means. In the physical world people collect books, papers, notes, physical objects. In the digital world people collect emails, documents, Web pages, pictures, videos, audios, etc. and a personal Information Management System (PIM) [15] is intended to support the user's activities by acquiring and managing (organization, maintenance, retrieval, sharing) their information for using it when needed. In this light, information comes in packages that encapsulates it, called *information objects* (a.k.a. information items). Such objects live in a common environment, a personal and unique *space of information*, and these should be organized in such a way that retrieving it in the right moment and for a concrete need will be possible. For such purpose, it is necessary to maintain the information objects mapped to different *needs*.

In the Web context, a PIM should offer the user the possibility of extracting information objects at one or many different granularity levels. For example, Evernote provides an extension called Web Clipper,[2] which allows end users to collect, annotate and tag full Web pages, but also some sections through HTML screenshots or DOM selection. Once collected, users can access the elements with their Web-based client, or the available native application for desktop or mobile devices. The client allows visualizing the list of harvested information objects, where users can edit, reorganize and search them by navigation or text-based search based on tags, annotations' content or the title.

In academic literature, we can also find tools supporting different approaches through Personal Information Managers taking information from the Web, like Piggy Bank [14]. In this work, the authors present a browser extension that lets users extract information objects from any Web page—regardless of the origin—and store them in RDF format. The ideal scenario for this approach is the one with an RDF-enabled Web site, but otherwise it could be possible by using screenscrapers to re-structure information into RDF. They contemplate different levels of granularity: the full page but also their components as information objects. Having all the information objects in the same format and space of information, it makes it possible for the items to be searched and organized taking advantage of the semantic information.

## 2.2 Web augmentation by end users

Web augmentation is a technique to adapt existing Web contents according to the requirements of individual end users [4]. It implies manipulating Web pages and providing an added value by changing some of their content, style or behaviour. In general terms, there are two strategies to implement Web augmentation: client-side or proxy-based. The difference among them is where the manipulation is carried out. In the first case, the code supporting the adaptation is attached to a Web page when the Web browser receives and parses its Document Object Model (DOM). In the second case, such responsibility resides on the proxy: when the client asks for a page, the proxy should request the original page (by an HTTP request to the server), get the response, apply the required changes and return such augmented version to the client. Indeed, in both cases, making these changes requires certain privileges to modify the structure, style or behaviour of a document. The former requires to extend the Web browser, which is a very common practice. The later requires to access to the proxy-based application to manipulate original content.

In this work, we focus on client-side Web augmentation, the most popular augmentation approach. An enormous number of browser extensions and userscripts[3] are published and updated every day, created to meet the need of adding certain functionality on the Web, either on a specific Web page or a set of them. For instance, The Camelizer[4] is a Firefox extension that enhances Amazon products with extra information by tracking the product's price history and comparing their prices against the ones offered by external shops. As a browser extension, it has the privileges to retrieve and include both, the external content and the produced information, in the same context of the current window.

When it comes to end users, it is possible to contemplate them as the creators of their own solutions through WA, although not with the same advantages as professional developers. First, because they do not have the same technical knowledge to design a good solution. Second, the end-user languages are usually not as flexible as the professional ones, because they tend to be shaped by constructs in a much higher level of abstraction than the ones usually offered in a professional development environment. Moreover, many end-user tools tend to allow the configuration of existing components rather than their composition for creating an application. Such difference is addressed in [18], as "Parameterisation or Customisation" against "Program Creation and

---

[2] Web Clipper by Evernote: https://evernote.com/webclipper/.

[3] Userscripts: https://userscripts-mirror.org/.

[4] The Camelizer: https://addons.mozilla.org/es/firefox/addon/the-camelizer-price-history-ch.

Modification". In this work, we focus on the second group, and we can devise two end-user roles in such a setting; the one of those who only use the applications others create, and the one of those who create an application by using any End-User Programming system [20]. The first is called *producer* and he is the one using end-user specialized tools that will empower him to create his own solutions. The second is the *consumer*, and he is just intended to use the existing solutions that other people has created (e.g. a producer or—why not—a *developer*).

It is not required for a producer to have technical background or know-how, but some of the tools require some code writing from end users, although in a much more abstract level than a professional programmer. For instance, Programming By Example [12] technique requires the end user to demonstrate how to solve a problem by interacting with the User Interface. There exist examples concerning the first group and implemented by extending the browser's capabilities [8, 17]. In both cases, the user's demonstrates how to solve a task and the involved steps are transparently recorded, so the system can reproduce them later. A similar approach that uses the Web pages as canvas is WebMakeUp [5].

Our approach is aimed for *producers* that do not have *programming skills*, but we also contemplate a more advance role: a *developer*. Note that our goal is not just to support a particular kind of DUI behaviour; moreover, we want to provide a platform for executing DUI behaviour while browsing Web sites. With this in mind, we conceive developers as the ones incorporating new DUI behaviour to the platform with, so the end user may choose among a wider set of constructs to use in their productions. Nevertheless, we provide several DUI behaviour already implemented in order to illustrate our approach.

## 3 Distributing existing Web user interfaces

The main idea behind the approach is to provide end users with a set of tools for defining their own DUI experiences according to their needs. Note that our aim is to support the distribution of Web applications under predefined designs (e.g. apply the same distribution every time the user accesses the same Web site from a desktop computer and a mobile device at the same time), but also we want support the specification of DUI under unexpected and opportunistic situations where the DUI layer was not previously defined (e.g. to quickly set a navigation control under an unplanned or sporadic use of a Web site).

To achieve such a goal, we rely on the adoption of a Personal Information Management (PIM) system that allows users to put together any UI component from all over the Web. The level of granularity of the collectable components is any DOM element, including the full page's body. In this context, elements from diverse Web applications coexist and are wrapped and interpreted as a collection of, what we called, UIObjects. Once in the PIM, the end user can apply DUI behaviours to the existing UIObjects on demand—to support opportunistic DUI requirements—or configure the automatic execution of DUI behaviours—to support frequent use. The last alternative is supported by a rule engine that observes UIObjects states and may bind DUI behaviours and UIObjects when an interaction event occurs.

Summarizing, we propose to contemplate both opportunistic and frequent needs by giving end users the possibility to:

– Select and collect UI elements into a PIM, from any Web site at any moment.
– Trigger DUI functionality for those UI elements on demand by interacting with elements collected into the PIM,
– Create rules for the automatic execution of such DUI functionality,

In this section, we present the full approach, its underlying concepts and architecture (in Sect. 3.1) and explains the two strategies for supporting end users to distribute elements from existing Web UIs (Sects. 3.2 and 3.3).

### 3.1 Concepts and architecture

We use the underlying philosophy of PIMs that allows users to put their relevant information objects into a common space of information, where such objects coexist and can be retrieved for interacting with them according to the purpose of the current user's activity. However, instead of collecting and structuring personal content such as traditional PIMs, we propose a UI-component-oriented PIM, where the user may select and harvest UI components from any Web site. In this setting, the information object traditionally contemplated by PIMs is, in our work, an UI object.

In the current work, we focus on the client-side features of the architecture, in which has three main artefacts:

– *UI Component* is the native UI component or widget that composes a particular Web site, for instance, a form, an image, a panel, etc. The main idea is that end users select the UI components they are interested in, so a UIObject is generated from a UI Component,
– *UI Object* this is a representation of a UI Component enriched with some added DUI capability. This behaviour is not associated in the original Web page, but through our tool. UI objects live in a UI-Component-oriented PIM,

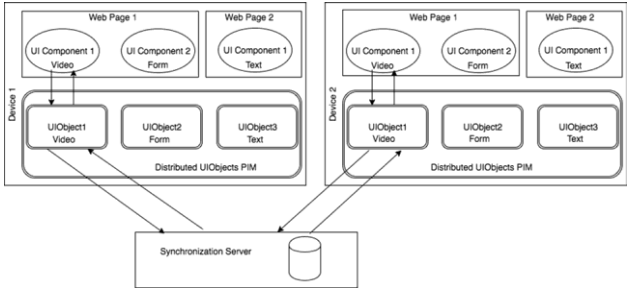

Fig. 1 Architecture schema of the UIObject-PIM platform

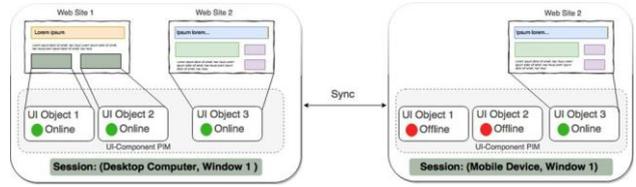

Fig. 3 General scheme of the PIM and the collected UIObjects

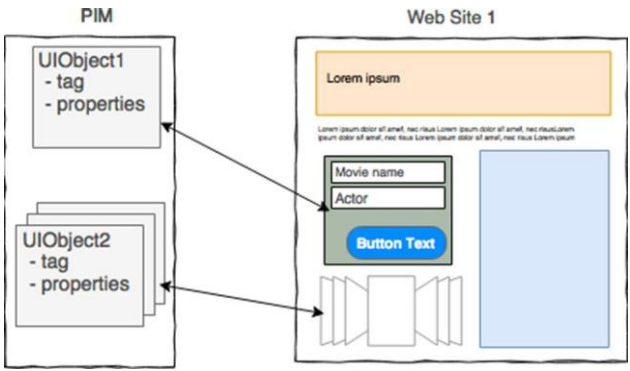

Fig. 2 Exiting UI components collected as UIObject

- *UI-Object-oriented PIM* is a PIM maintaining references to a collection of UIObjects created by the end user. It requires authentication, so the UIObjects collected by the end user can be synchronized and retrieved from any of his devices. The UIObjects collected into the PIM decorate UI components with specific DUI-based behaviour. Such behaviour is materialized for each UIObject as operations listed in a contextual menu. For example, to render a UI Component only in one of the user's devices, he must run the "Show only in..." operation, that will ask the user which for a target device and then to carry on the desired UI effect.

Most of the approach works at client-side. As shown in Fig. 1, the UIObject-PIM is implemented as a Web Browser extension that allows end users to share a unique space of information among their own devices. Users can install multiple UIObject-PIM in their devices, and the contained elements are synchronized through a synchronization server. Each UIObject has an internal state in a PIM, allowing the system to know in which devices the UI component is active. The approach is not constrained to a centralized server; the PIM may be configured to work with any specific deployment of the application server.

Figure 2 shows how a form or any other DOM element is wrapped by UIobjects and represented in our PIM.

In our approach, we offer different ways to select and collect UI objects into the PIM. For supporting this, UIObjects are aware of the type of DOM element being wrapped. For instance, in the Fig. 2, *UIObject1* must know that the DOM element referenced by the UIComponent is a form. Meanwhile, *UIObject2* is aware that it manages a collection of images (img HTML nodes). This harvesting mechanism allows users to collect arbitrary DOM elements from existing Web sites. UIObjects are not necessarily static, i.e. these are not a copy of the DOM element but a wrapper that dynamically obtains the corresponding DOM element.

As a DOM element represents a particular UI element, these elements could be very specific (for instance a form) or comprehensive (a full Web page). For example, the end user may want to distribute the tendencies box of Twitter to a second session for remote control, or he may collect the full Web page for visualizing all the images found in the full body of a blog entry into a second session.

At this point, we want to make some important highlights about the PIM and the UIObjects:

1. The PIM is deployed as a Web browser extension synchronized through a Web application. It is a distributed PIM; the same user may use different devices and, therefore, different instances of the client should be installed in the browsers,
2. DUI augmentations are performed at client-side,
3. All user devices will maintain the same UIObjects into the PIM, without taking into account which device had been used for collecting them.
4. Different DUI features can be applied to the same UIObject. These DUI features are implemented as DUIBehaviours, which are extensible for developers and available for end users, so they can easily use them,
5. Each UIObject has an internal state (online/offline) for each opened PIM session. This state indicates if the DOM element being represented by the UIObject is currently visible in the Web page that the user is visiting in that session,
6. PIM sessions share the state of their UIObjects instances.

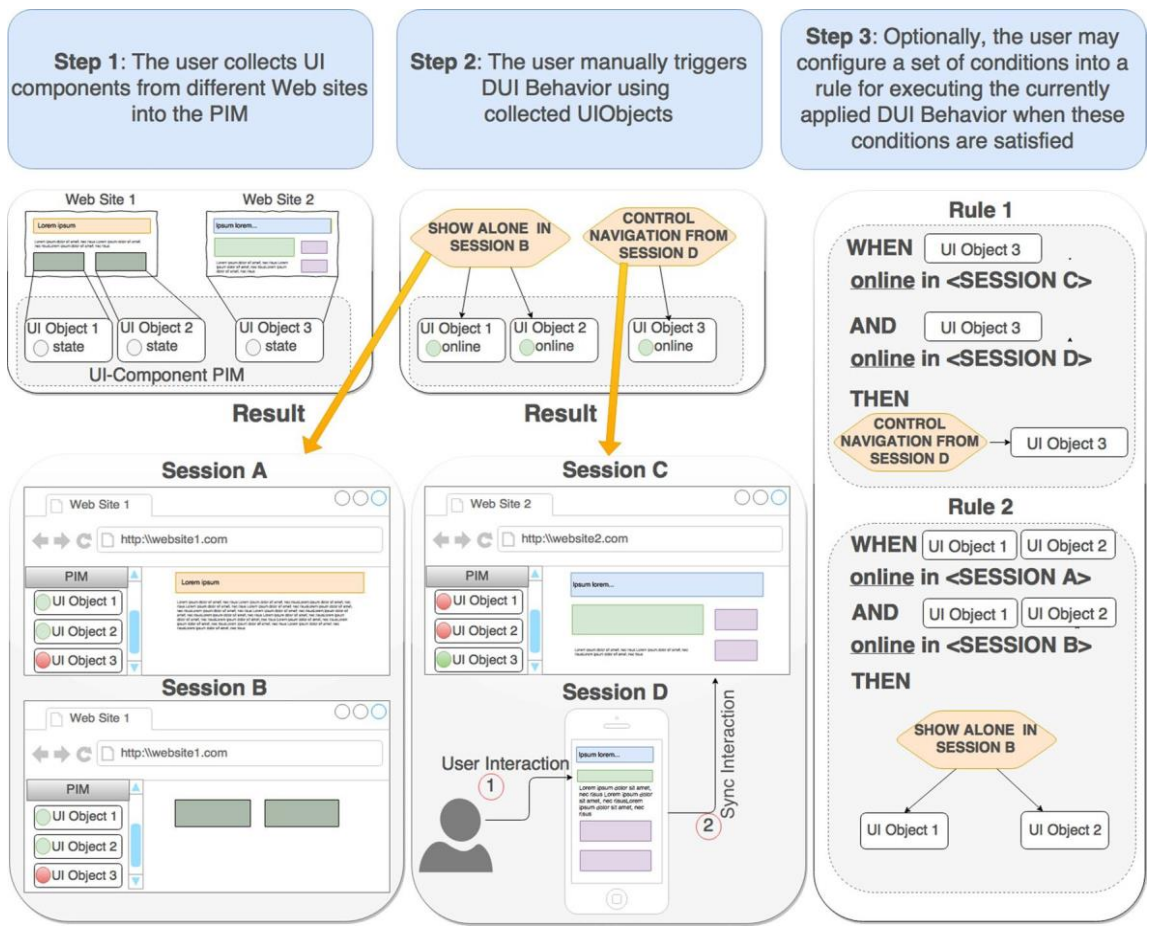

Fig. 4 Manually DUI behaviour execution (step 2) and automatic DUI execution (step 3) based on collected UIObjects (step 1)

These points are illustrated in Fig. 3. In this schematic graphic representation, two users sessions opened in different devices, a desktop computer at the left and a mobile device at the right. First, note that the same UIObjects are shown in both sessions; however, for each session UIObjects have different states. For instance, given that from the mobile session the only Web site in use is *Web Site 2*, then only UIObject appearing as online is *UIObject 3*, in the opposite case, *UIObject 1* and *UIObject 2* appear as offline. Each PIM session is aware of the state of all UIObjects in the device where it is running; however, it can also request such state from other PIM session in use.

Once into the PIM, UIObjects may be used to trigger DUI behaviours. A DUI behaviour is a JavaScript program (available in all user sessions) that is executed for a specific UIObject. For instance, the DUI behaviour "Redirect interaction" captures the UI events generated by the user's interaction with a DOM element (which is wrapped by the UIObject) in the session where they were originated and redirect it to another session (i.e. to reproduce the same events). The behaviours are offered to the user as menu options in the PIM, as it is shown in the following sections. Although we have developed some of these DUI behaviours for illustrating our ideas, actually our approach is flexible in this way given that new behaviours may be developed (by developers) and installed (by end users).

DUI behaviours may be executed directly by the user. For instance, the user may activate the DUI behaviour "Show alone in..." for a UIObject. This action would hide that UIObject in all opened sessions, and it would appear only in the selected session. This simple behaviour would imply to execute some functionality in all the devices that are currently in use. This is depicted in Fig. 4, specifically on step 2, where two behaviours are manually applied by the user. The first, "Show alone in on Session B", makes possible to distribute UI components on different displays using the same device (a desktop computer). The second, "Control Navigation from Session D", allows the user to control the interaction in a Session (C) from another device (Session D).

These two examples (Fig. 4, Step 2) are intended to satisfy opportunistic needs about a distributed use of the application. However, it would not be very useful if the user has to trigger the desired DUI behaviours every time he wants a distributed experience. In cases like this, where frequent scenarios of use exist, our approach supports users with the possibility of defining execution-context rules.

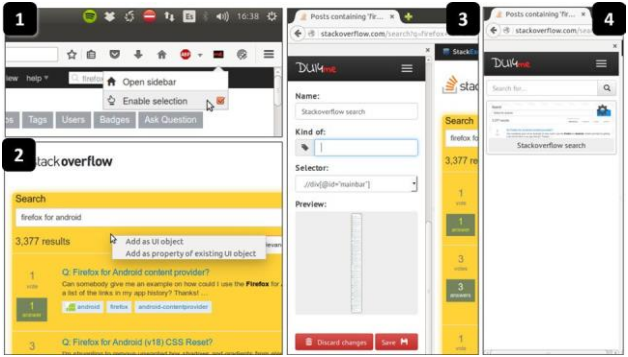

Fig. 5 Defining a UIObject

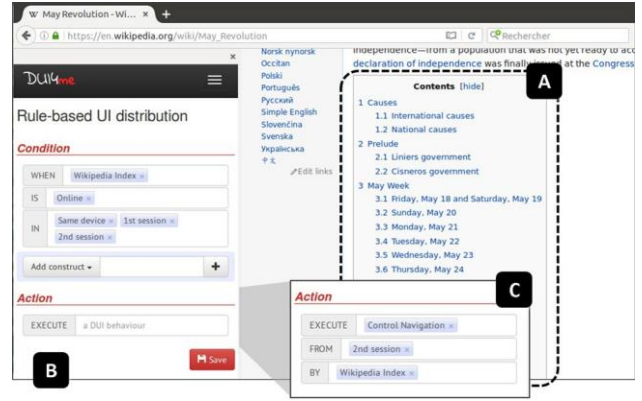

Fig. 6 Defining a distribution rule

Together with the PIM, there is a rule engine that deals with different aspects of UIObjects, among them, it knows the internal state of UIObjects. In this way, the rules may be aware of UIObjects state. Consider the example in Step 3 of Fig. 4; when the UIObject's state changes to online in a concrete session, a DUI behaviour could be automatically triggered. Therefore, the same behaviour manually triggered in Step 2, is now configured to be executed automatically under certain conditions.

By the creation of rules, users may specify how to distribute a Web site for frequent uses. For instance, in the case of Rule 2, when the UIObject 3 is online in Session C (a desktop session) and in Session D (a mobile one) then the "Navigation control" is triggered. Note that UIObject 3 represents the entire body element of the Web site 2's DOM.

The remaining of this section introduces the underlying architecture (Sect. 3.1), shows how users may collect UIObjects and trigger behaviour manually (Sect. 3.2), and finally explains how users may configure rules (Sect. 3.3).

### 3.2 UIObjects harvesting and manual distribution

A UIObject is a JavaScript object abstracting a UIComponent that the user has selected. In Fig. 5, a user is enabling the DOM selection in step (1), so he can highlight the DOM elements for selection, as in step (2), with a proper context menu enabling the harvesting. Once selected, he can configure some properties, such as the name and the tag shown in (3). The most relevant one is the UIComponentStereotype. It allows users to choose the kind of UI widget, such as image, text, form, video, etc. Then, all UIObjects are available in the default view of the PIM, as shown in (4), so the user can interact with them through the offered behaviour.

When a UIObject is collected into the PIM, DUI-based behaviours are attached to it under the basis of the decorator pattern [10]. There are two kinds of DUI-based behaviours: stereotype-agnostic, that could be attached to every UIObject, since their goals are compatible with all kind of UIComponents; and stereotype-specific, which are attached only to those UIObjects collected as a specific UI stereotype. An example of the last case could be a YouTube video, which can offer more specialized operations than the base components, such as "Play video on..." a concrete device.

It is important to note that the set of decorators that are applied to a UIObject can be configured by the end user, by clicking the gear icon shown in step (4) of Fig. 5. Moreover, although we provide some basic behaviours, new decorators may be defined by developers and imported into the platform, as explained in Sect. 4.3. Extending a behaviour requires programming with client-side Web technologies (HTML, JavaScript, CSS), specially using a JavaScript API that allows developers to manipulate the UIObjects collected into PIM.

Extending the existing behaviours allow developers to support new kinds of DUI applications, like a particular kind of distributed layout, but also new specific operations, like a specialized "play" and "stop" messages for the YouTube player. If such DUI behaviour is imported into the PIM, when a user collects an object into the PIM and apply such decorator, such messages could be sent transparently from any device.

### 3.3 Rule-based UI distribution

There is a second strategy for applying a DUI behaviour to the existing harvested UIObjects, which is by creating rules that support the frequent use of the augmented distributions. Consider a user that usually navigates the Wikipedia web site and wants to distribute the index of every article to a second session, so he can control the navigation from a second window. He has already collected the index of an article in the PIM, the one marked in Fig. 6 as (A), so now he wants to define the rule.

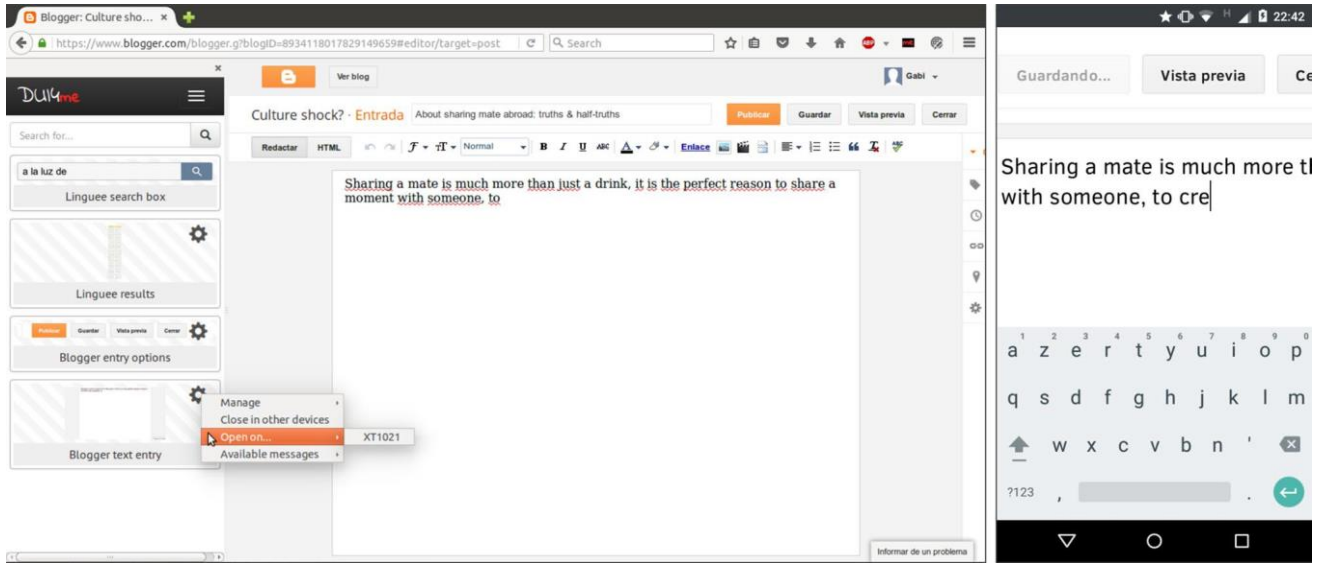

Fig. 7 Distributing UI components from a Web application to a mobile session

In (B), the user has defined the condition of the rule. It states that the "Wikipedia Index" will be distributed to the second section of the same user's device, when such object is online in both sessions of the same device. Being "online" means that the URL (or pattern) associated to the UIObject is opened, and that an instance of the wrapped DOM element could be found in its body. The user can attach extra declarations to the rule, by selecting a new construct at the bottom of the Condition section. Then, some possible target elements are shown through autofill while he writes, and once he selects one, it is rendered as a tag. He can also change the order of the statements of the rule by drag and drop.

Finally, the user should specify the action. Such section is dynamically loaded according to the selected DUI behaviour, as shown in (C). The "Control Navigation" DUI, in order to properly enable the remote control, it requires to know the UIObject in charge of controlling the navigation and the target session to distribute it. In this case there is just one associated behaviour, but the user can define more than one for being executed when the same condition of the rule is meet. This way, the user has a wide range of possibilities and total freedom to define his own rules.

## 4 Case studies and prototype

Although the approach lets developers to create new DUI behaviours, we have developed a set of predefined ones. So far we have identified: opportunistic and volatile interface migration, multi-monitor layout, distributed layout, messaged-based remote control, navigation control and remote UI Effect.

We have implemented two prototype client-side tools supporting the approach. The first is a Firefox 44 extension for desktop edition. This extension is the only software artefact that users need to install, which already supports several DUI behaviours by default. However, since we do not want to limit the tool to the default DUIBehaviours, the tool allows to install further ones (those created by developers) from the repository.

The second is a Firefox for Android extension, for version 42.0a1. At server-side, we implemented a minimal functionality to synchronize changes of the UIObjects both tools manage.

In this section, we illustrate two case studies; the first focused in opportunistic distribution, and the second in frequent distribution through the definition of predefined rules.

### 4.1 Using UI-PIM for opportunistic DUI requirements

Consider a student, Máximo, who is living abroad for some months and he wants to register his experience in a blog. Sometimes he can write a long time from his desktop computer, but there are situations in which he migrate the task to his mobile device. He is writing the blog in English but his mother tongue is Spanish, and sometimes he has some doubts about language expressions that he uses to checks them out in linguee.com. He would like to distribute some elements of the Linguee and Blogger Web sites in his mobile to make easier this task.

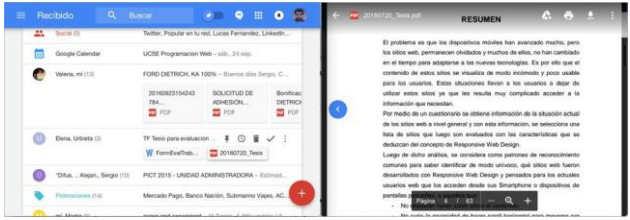

Fig. 8 When the user clicks on a GmailAttachment UIobject in Session A, the corresponding interaction is triggered in Session B

In Fig. 7, we show our tool's sidebar with some UIObjects created for the mentioned scenario; a search box and a results box from Linguee, and an entry options and a text area from Blogger. With such objects in the PIM, Máximo can open such elements in his mobile (the *XT1021* listed device), as shown in the left of Fig. 7. If triggered, such message will notify the mobile browser PIM to open the same URL but just to open the defined UIObjects the user has defined from the desktop PIM.

Blogger does not provide a mobile version, is not responsive, and does not offer an up-to-date sync functionality among different devices. With our tool, while Máximo changes the blog entry's document, every modification is notified to our server, who asks every registered listener (other user's PIMs) to update the view of the element in the proper context. This way, the functionality of the site is not altered by our adaptation; the automatic saving functionality by Blogger being executed is shown at the right screen of Fig. 7.

### 4.2 Supporting frequent DUI requirements using UI-PIM and Rules

Two simple DUI examples based on rules are explained in this section. In Fig. 8, we show an example for applying distributing interaction on Google Inbox when it is used from a single desktop computer with two displays. In this case, the collected UI components are related to the attached files from emails. When that UIObject is online in both sessions (a session per display), the DUI behaviour "Redirect interaction to..." is applied. In this way, when the user clicks on one of the attached files, this is opened in the second session in order to maintain visible the list emails in the first one.

A second example, where the scenario of use involves a desktop computer and a mobile is defined for Vimeo.com, is shown in Fig. 9. Here, the entire body element (DOM element) is collected as a UIObject and the "Control navigation from..." behaviour is triggered with this object in order to command the Session B's UI from the mobile device.

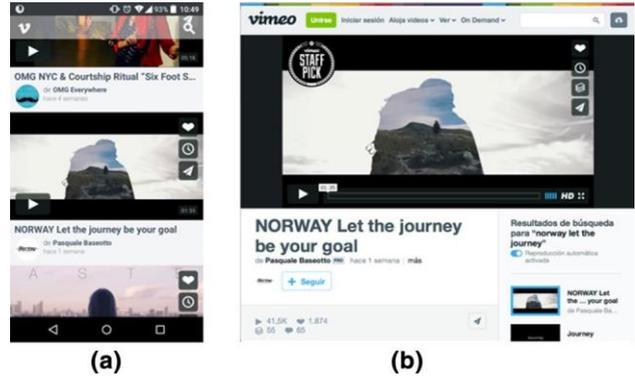

Fig. 9 When the user navigates from the mobile Session A, the navigation effect is actually executed in the desktop Session B

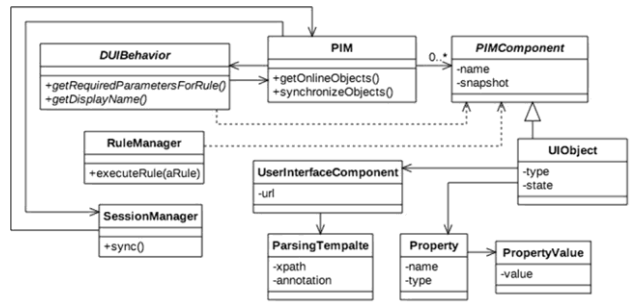

Fig. 10 DUI behaviour code excerpt

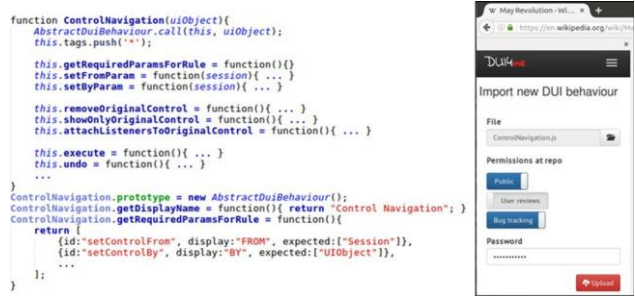

Fig. 11 DUI behaviour code excerpt

### 4.3 DUI behaviour framework

In order to support new DUI behaviours, we allow developers to extend the framework with their own classes. A simplified UML class diagram of this software is shown in Fig. 10. Specifically, to create a new DUI Behaviour, developers need to create the new class by extending the AbstractDuiBehaviour or any of its subclasses. As it is shown in Fig. 11, it is mandatory to implement two class methods for helping our tool to display the parameters required by the class before properly instantiating it. For instance, the

(C) point in Fig. 6 presents the required parameters by the ControlNavigation DUI behaviour. Then, developers should implement the right setters for such parameters (e.g. setControlsFrom, setControlsBy), so the authoring tool can properly instantiate the class.

Then, developers can implement the desired DUI behaviour through their own methods, like removeOriginalControl() and showOnlyOriginalControl(). According to the given example in Sect. 4.2, the first method will be called in the first session, and the second method in the second one. This way, the controlling UIObject (e.g. the one wrapping an index) is hidden in the first session and the remaining elements of the Web page are hidden in the second session. Moreover, in this example, there is a method in charge of adding a listener to the DOM elements of the observed UIObject (the one in the second session), so they can post the proper sync messages to the server. The DUI behaviours should implement—at least—as many distributions as sessions they want to support. Execute will receive the right parameters (indirectly from the emitter through the sync server), so it can choose the right method to call.

Finally, the extended class can be imported to be used by the PIM, as shown in the right side of Fig. 11. Users can choose to upload the file for personal use or to share and store it into the public repository, so other users can make use of it. From the PIM, there are two general configurations the owner of a DUI Behaviour can choose: to enable user reviews and bug tracking. However, a broader set of options is available in the Web site of the repository.

### 4.4 Sessions and synchronization

There are some challenges regarding how manage different sessions and how the synchronization is made. We want to support DUI even when the user is accessing Web content from a single computer device (for instance a desktop computer with two displays). With this in mind, a session is not limited to a device, but to every Web content display. Synchronization among sessions is necessarily made through the server-side application, i.e. using Web Sockets. In future implementations of our prototype, we will investigate how WebRTC may be used to synchronized sessions among them without using server-side application.

## 5 Related works

There are several frameworks facilitating the creation of Web applications with DUI features. CTAT [3] supports the development of cross-device Web applications using touch and tilt—motion-based—interactions. It is designed for rapid prototyping development, it supports many-to-many communications, cooperative interactions, and developers can specify concrete interactions for four kinds of devices: laptop, tablets, smartphones and smartwatches. Connichiwa [22] is a framework for creating cross-device Web applications. It allows developers to implement their applications with no external server dependency; applications built with this framework run as a local Web application and a native helper runs a Web server on demand, when a session is created, in one of the participating devices. Although they mention the non-dependency of the user's position and they take into account the arrangement of devices when sharing a session, in this work, there is no mention about the use of context information to automatically trigger a distribution. Panelrama [26] is a framework allowing developers to specify DUI features in their Web applications, by providing a new XML panel element, where DOM elements should reside in order to be distributable. In contrast with our work, it is possible to distribute applications interactions from scratch through such frameworks; however, to use it with an already existent application involves restructuring it or, in case of the third-party ones, it is not possible. Moreover, such approaches are directed to developers, not giving end users the chance to define the distributable components of the application.

WebSplitter [13] is also a framework for distributing user interfaces, although with the particularity that it is not among devices, but also among different groups of users. Authors should specify an XML-based Web application with no tag type restrictions, but uniquely identifying each one. Then, it is required also an XSL-specification for the usage of each tag and, finally, an optional policy file indicating which XML tags should be distributed to which group of users and kind of device. A proxy is in charge of splitting the document, establishing the browsing session and controlling the collaboration between devices. In contrast to our approach, this one requires the end user to have some programming knowledge, and it is not about reusing existing Web content or keeping and combining the user's UI objects of interest in a common ambient.

An approach intended to support the authoring of distributable user interfaces with cross-device interactions features is presented in [21]. In this work, the authors propose to implement a model and a framework supporting the rapid prototyping of applications with DUI features by developers, and an authoring tool for supporting the distribution of their elements and interactions by end users. Conceived in the foundation of active components, users will be able to share UI elements and each of them will trigger a single action. They also plan to share a combination of such components and actions in a package. A similar approach is presented in [19], where developers create the applications with an initial DUI configuration and the end users can re-configure the distributions at runtime with a customization tool. Both approaches are similar to ours in the sense that they contemplate the end users improving their experiences over existing user interfaces; however, they contemplate that for applications built with their framework (no matter if it is a third-party one); meanwhile, we support it for all Web, by taking any Web application as a base for applying an augmentation layer.

In [24] the authors present an interface distribution daemon that, based on service descriptions (made in RelaxNG), distributes the components of an application in three ways: user-driven, system-driven and continuous. In all cases, such distribution is performed among the devices that are registered and active in a personal interaction space. In the first modality, the user is in charge of choosing which services he wants to distribute in which device. In the second one, such decisions are taken by the daemon, based on a cost function that analyses the weight of the service according to the screen space. In the last case, decisions are taken according to changes in the interaction space, that is, devices entering and leaving it. This approach does not consider any third-party existing Web site as potential target for the distribution, and the end user's participation is limited to the use of an already specified distributable application; they do not take decisions about what elements of the UI are distributable or not. Our approach is also different regarding the storage and the access of the distributed elements in a shared space of information.

Other works face distribution through migration. [9] presents a framework and runtime support for allowing developers to create applications supporting DUI with multi-device and multi-user support. However, it does not consider the development by end users or the support of experiences that combine elements of different applications. The research group of [1] presents a model-based approach supporting total and partial migration of user interfaces. They rely mainly in applications created with a tool named TERESA, which produces a description about Web pages and the interactions they support. Such description is used to split the UI, to decide which elements can be migrated to a target device (e.g. because they have no dependencies to another element). Adaptation is performed in the UI if partial migration is chosen, and usability issues are taken into account. This environment could be also used with Web interfaces did not created with TERESA, but due to the lack of the underlying description, it cannot benefit from the adaptation mechanisms. Different to our approach, the context values cannot be used for exploring the distribution of user interaction across diverse applications.

Below, we present some works where the end user role changes, from a simple consumer to the one who produces— by creating or configuring—the distribution of interactions of an interface. Such works also represent another strategy for distributing user interactions, concerning the manipulation of existing Web applications without altering source code, i.e. building a solution based on existing ones.

In contrast to the last mentioned work and in line with our approach, Proxywork [25] allows users to distribute the user interface components of any Web application—even not designed for this purpose—in different platforms. It does not depend on native components, but in a proxy that manipulates the requested Web pages, injecting it extra code in order to add the distribution operations and to be aware of the existence of other possible target devices to perform the distribution. A main technical difference regarding the augmentation technique is that our approach works at client side. Other bifurcations are context awareness and migration features. In [11], the authors contemplate distribution through the partial migration of certain portion of a Web application's interface. It is intended for empowering the end user to opportunistically select the elements he is interested in and migrate them from a desktop environment to a mobile one. For this purpose, the devices involved in the interface migration must run an application for making them aware of the existence of each other, and make them selectable for the migration. Web pages are intercepted by a proxy and some annotations are added before being delivered to the Web Browser. When the user chooses to migrate the current Web page, a reverse engineering module analyses the Webpage at proxy-side, and produces a description, which is received by an orchestrator and a list of elements arranged hierarchically is sent and presented into the client so the user can select which of them to migrate. In contrast to our approach, in this work, the authors do not consider the possibility of combining UI components of multiple Web applications, nor the use of the context information.

As shown in Table 1, all the mentioned works support both, input and output redirection, and most are oriented to create applications from scratch, or requires the manipulation of the existent application. At least 63% of the works implement migration. An 18% allow end users (with no programming skills) to choose which items are distributable in the Web app interface, no matter if the application was created with their own framework or if it is a third-party one. Another 18% takes into account the user's context information to automatically distribute the UI. In this category, we do not consider the ones that involve the user's interaction, even if such interaction is decoded by context information (e.g. the orientation supporting tilt gestures). None of the works is aimed at presenting distributed components from diverse applications, nor to keep them out of its context, in a dedicated and personal space of information.

In this work, we propose to empower end users with the capability to apply an augmentation layer with DUI behaviour over existing Web pages. In contrast with existing works, we do not provide an end-user tool for executing already defined distributable applications but for enabling them to choose which elements they want to distribute, and improve an existing Web interface. Note that the DUI features in our approach are not predefined, since we provide a framework for developing new DUI behaviour and a PIM-based architecture to execute that behaviour to DOM elements selected by end users. Thanks to the possibility to define any element—even the whole Web page—as a distributable component, wrapping it with extra and specialized functionality, our approach may support several kinds of DUI dimensions, such as light interface migration, remote control, distributed layout, etc. It is important to mention we support two DUI scenarios, one based on opportunistic needs and another based on recurrent uses.

Table 1 Features contemplated by the analysed works

| | Our approach | [1] | [3] | [9] | [11] | [13] | [19] | [21] | [22] | [24] | [25] | [26] |
|---|---|---|---|---|---|---|---|---|---|---|---|---|
| Input and output redirection | ✓ | ✓ | ✓ | | | | | | | | ✓ | ✓ |
| Migration | ✓ | ✓ | | ✓ | ✓ | ✓ | ✓ | | | ✓ | | ✓ |
| Automatic distribution under certain context values | ✓ | | | | | | | | ✓ | ✓ | | |
| Storage of distributed elements in a global space of information | ✓ | | | | | | | | | | | |
| Combination of UI elements of diverse Web applications | ✓ | | | | ✓ | | | | | | | |
| Distribution of any element of any Web app (including third-party ones) | ✓ | | | ✓ | | ✓ | ✓ | ✓ | ✓ | ✓ | ✓ | ✓ |
| From scratch or requires altering the existing sources | | | | | | | | | | | | |

## 6 Discussion

This paper addresses the problem of DUI in the context of existing Web applications. From our point of view, several kinds of user interactions that may be supported directly from Web applications could also be seen as part of the web browsing activity. Although we are aware of existing works supporting developers in the development of Web applications and offering DUI experiences, we have showed that DUI could be addressed not only from the Web application, but directly from the Web browser. This moves the discussion from DUI Web applications to DUI Web browsing. Web augmentation (also known as Augmented Web Browsing) gives the possibility to end users to adapt how they interact with Web applications. This technique is perfectly suitable with the needs of distributed user interaction. In this way, supporting DUI directly from Web Browsers allows users to satisfy by themselves their DUI requirements. As a first consequence, end users would drive how an application may be distributed (in terms of UI and interaction), but also would be able to do it with any existing Web applications, beyond how these applications either support or not support DUI features natively.

A second consequence of enabling Web Browsing DUI is the falling of application boundaries, i.e. instead of only provide a DUI experience in the context of a single Web application (distributing UI components belonging to it), we will be able to study how cross-application DUI may improve user experience. For instance, when visiting two Web applications that are related to a particular task (such as booking hotel rooms in one application while looking for flights to the city in another one) the interaction made in one of them could impact on the other application by form filling, highlighting related information, etc. Augmentation-based DUI widgets could be explored. For instance, when reading a Wikipedia article in a browser window, in a second one the user could show a carousel with images from Google images. Similar examples could be defined in other domains.

For instance, imagine a journalist visiting different news web portals. In this scenario, when the user is interested in a particular topic in a portal A, then, related news from the second portal could be focused, highlighted, etc.

As a final comment, we believe that DUI requirements go from opportunistic uses to predefined designs that are use recurrently. Our approach aims to support with a single architecture these two kinds of needs.

## 7 Conclusions and future works

In this paper we put together two trends related to the HCI in the context of the Web: distributed user interfaces and Web augmentation. The first improves user experience by considering the complex and current interaction with Web applications, which could occur at the same time from different devices. The second lets users customize (by using end-user development tools) how Web applications are perceived, in terms of content and functionality. We envisioned then the concept of Distributed Web Browsing, allowing users to specify when and how Web applications are distributed, even when these applications do not support DUI natively. Our approach is not focused on giving a predefined set of DUI functionality to users. Instead, we provide a framework to allow advanced users (those with programming skills) to create new DUI functionality (called DUI behaviours) that could be executed to specific DOM elements. These DOM elements are specified by end users and collected into a UI-oriented PIM. Users may trigger DUI behaviour manually. At the same time, the use of a PIM allows the execution of rules that apply the desired DUI behaviour when the conditions are satisfied. In this sense, non-opportunistic DUI requirements are supported.

Summarizing, we designed a flexible architecture based on a UI-object-oriented PIM, which was coined with several premises in mind:

1. Users should be able to collect UI elements as UIObjects into the PIM,
2. Every UIObject has an internal state in each session opened by the user,
3. Users should be able to trigger DUI behaviours directly to UIObjects from the PIM, in order to support opportunistic requirements,
4. Users should be able to specify the automatic execution of DUI Behaviour by defining rules. A rule may involve more than one condition and more than one behaviour,
5. Developers should be able to create new kinds of DUI behaviours compatible with existing UIObjects, i.e. distributing the UI components in new ways, no matter the underlying Web sites DOM structures, i.e. allowing the reuse of behaviours in several Web applications,
6. DUI Behaviours, developed by end users with programming skills, have access to the UIObjects, their states and intrinsic properties and behaviour.

Considering that at this stage of the project, we were focused in demonstrating the technical feasibility of the approach, we will explore more technical details that can improve the performance of the user experience. For instance, a following step in the agenda is to test other inter-PIM communication mechanisms. Likewise, due to the constantly changing nature of the Web, we are also looking for alternatives to provide wrappers with resilience capability, so that when the structure of a Web site changes, the user experience is affected as little as possible.

We are currently designing an experiment to demonstrate the operational feasibility of the approach. The main idea is to evaluate the approach with real end users with no programming skills in order to validate the operational feasibility and measure the potential user adoption rate.

After fulfilling that goals, we will continue this research by analysing how cross-application DUI and augmentation-based DUI could improve user experience and in which ways this could be applied using our approach. With this in mind, we plan to take advantage of the possibility for adding semantic tags to UIObjects that the platform gives when the user collects UI elements.

Acknowledgements This work was supported by STIC AMSUD project WAMAW-OUR: Web Augmentation Methods for Adapting Web Sites for Supporting Opportunistic User Require-ents

## References

1. Bandelloni, R., Paternò, F.: Flexible interface migration. In: Proceedings of the 9th International Conference on Intelligent User Interfaces, pp. 148–155. ACM (2004)
2. Bharat, K.A., Cardelli, L.: Migratory applications. In: Proceedings of the 8th Annual ACM Symposium on User Interface and Software Technology, pp. 132–142. ACM (1995)
3. Di Geronimo, L., Husmann, M., Patel, A., Tuerk, C., Norrie, M.C.: Ctat: tilt-and-tap across devices. In: International Conference on Web Engineering, pp. 96–113. Springer (2016)
4. Díaz, O., Arellano, C.: The augmented web: rationales, opportunities, and challenges on browser-side transcoding. ACM Trans. Web (TWEB) **9**(2), 8 (2015)
5. Díaz, O., Aldalur, I., Arellano, C., Medina, H., Firmenich, S.: Web mashups with WebMakeup. In: Daniel, F., Pautasso, C. (eds.) Rapid Mashup Development Tools. Communications in Computer and Information Science, pp. 82–97. Springer, Cham (2016)
6. Elmqvist, N.: Distributed user interfaces: state of the art. In: Gallud, J., Tesoriero, R., Penichet, V. (eds.) Distributed User Interfaces, pp. 1–12. Springer, London (2011)
7. Firmenich, S., Bosetti, G., Rossi, G., Winckler, M.: Flexible distribution of existing web interfaces: an architecture involving developers and end-users. In: 5th Workshop on Distributed User Interfaces, ICWE (2016, in press)

8. Firmenich, S., Rossi, G., Winckler, M.: A domain specific language for orchestrating user tasks whilst navigation web sites. In: International Conference on Web Engineering, pp. 224–232. Springer, Berlin, Heidelberg (2013)
9. Frosini, L., Paternò, F.: User interface distribution in multi-device and multi-user environments with dynamically migrating engines. In: Proceedings of the 2014 ACM SIGCHI Symposium on Engineering Interactive Computing Systems, pp. 55–64. ACM (2014)
10. Gamma, E., Helm, R., Johnson, R., Vlissides, J.: Design Patterns: Elements of Reusable Object-Oriented Software. Pearson Education, London (1994)
11. Ghiani, G., Paternò, F., Santoro, C.: On-demand cross-device interface components migration. In: Proceedings of the 12th International Conference on Human Computer Interaction with Mobile Devices and Services, pp. 299–308. ACM (2010)
12. Halbert, D.C.: Programming by example. Doctoral dissertation, University of California, Berkeley (1984)
13. Han, R., Perret, V., Naghshineh, M.: WebSplitter: a unified XML framework for multi-device collaborative Web browsing. In: Proceedings of the 2000 ACM Conference on Computer Supported Cooperative Work, pp. 221–230. ACM (2000)
14. Huynh, D., Mazzocchi, S., Karger, D.: Piggy bank: experience the semantic web inside your web browser. In: International Semantic Web Conference, pp. 413–430. Springer, Berlin, Heidelberg (2005)
15. Jones, W.: Personal information management. Ann. Rev. Inf. Sci. Technol. **41**(1), 453–504 (2007)
16. Kim, K., Javed, W., Williams, C., Elmqvist, N., Irani, P.: Hugin: a framework for awareness and coordination in mixed-presence collaborative information visualization. In: ACM International Conference on Interactive Tabletops and Surfaces, pp. 231–240. ACM (2010)
17. Leshed, G., Haber, E.M., Matthews, T., Lau, T.: CoScripter: automating and sharing how-to knowledge in the enterprise. In: Proceedings of the SIGCHI Conference on Human Factors in Computing Systems, pp. 1719–1728. ACM (2008)
18. Lieberman, H., Paternò, F., Klann, M., Wulf, V.: End-User Development: An Emerging Paradigm, pp. 1–8. Springer, Dordrecht (2006)
19. Manca, M., Paternò, F.: Customizable dynamic user interface distribution. In: Proceedings of the 8th ACM SIGCHI Symposium on Engineering Interactive Computing Systems, pp. 27–37. ACM (2016)
20. Nardi, B.A.: A Small Matter of Programming: Perspectives on End User Computing. MIT press, Cambridge (1993)
21. Sanctorum, A., Signer, B.: Towards user-defined cross-device interaction. In: 5th Workshop on Distributed User Interfaces, ICWE (2016, in press)
22. Schreiner, M., Rädle, R., Jetter, H.C., Reiterer, H.: Connichiwa: a frame-work for cross-device web applications. In: Proceedings of the 33rd ACM Conference Extended Abstracts on Human Factors in Computing Systems, pp. 2163–2168. ACM (2015)
23. Vanderdonckt, J.: Distributed user interfaces: how to distribute user interface elements across users, platforms, and environments. In: Proceedings of XI Interacción, p. 20 (2010)
24. Vandervelpen, C., Vanderhulst, G., Luyten, K., Coninx, K.: Lightweight distributed web interfaces: preparing the web for heterogeneous environments. In: Web Engineering, pp. 197–202. Springer, Berlin, Heidelberg (2005)
25. Villanueva, P.G., Tesoriero, R., Gallud, J.A.: Proxywork: distributing user interface components of web applications. In: DUI@EICS, pp. 58–61 (2013)
26. Yang, J., Wigdor, D.: Panelrama: enabling easy specification of cross-device web applications. In: Proceedings of the 32nd Annual ACM Conference on Human Factors in Computing Systems, pp. 2783–2792. ACM (2014)